
\input phyzzx

\nopagenumbers
\line{\hfil CU-TP-605}
\line{\hfil hep-ph/9308054 }
\vglue .5in
\centerline {\twelvebf 	MONOPOLE VECTOR SPHERICAL HARMONICS}
\vskip .3in
\centerline{\it Erick J. Weinberg}
\vskip .1in
\centerline{Physics Department, Columbia University}
\centerline{New York, New York 10027}
\vskip .4in
\baselineskip=14pt
\overfullrule=0pt
\centerline {\bf Abstract}

Eigenfunctions of total angular momentum for a charged vector field
interacting with a magnetic monopole are constructed and their
properties studied.  In general, these eigenfunctions can be obtained
by applying vector operators to the monopole spherical harmonics in a
manner similar to that often used for the construction of the ordinary
vector spherical harmonics.  This construction fails for the harmonics
with the minimum allowed angular momentum.  These latter form a set of
vector fields with vanishing covariant curl and covariant divergence,
whose number can be determined by an index theorem.

\vskip .1in
\noindent\footnote{}{\twelvepoint \noindent This work was supported in part by
the US Department of Energy.}

\vfill\eject

\baselineskip=20pt
\pagenumbers
\pageno=1

\def\refmark#1{~[#1]}

\def\hr{{\bf \hat r}}
\def\c{{\bf C}}
\def\cl{{\bf C}^{(\lambda)}}
\def\clp{{\bf C}^{(\lambda')}}
\def\cls{{\bf C}^{(\lambda)*}}
\def\tD{\tilde D}
\def\y{{\bf Y}^{(q)}_{JLM}}

\def\calD{{\cal D}}
\def\calO{{\cal O}}
\def\u{{\bf U}}

\chapter{Introduction}

When analyzing fields in a spherically symmetric background, it is
often useful to expand the field in eigenfunctions of angular
momentum.  In most cases these are simply the spherical harmonics for
scalar fields, while for fields of higher spin one can construct
spinor, vector, or higher  harmonics.  The situation is somewhat
more complicated when one considers charged fields in the background
of a magnetic monopole.  Superficially, this is because the
electromagnetic vector potential is not manifestly spherically
symmetric, even though the corresponding magnetic field is.  At a
deeper level, it is a consequence of the extra angular momentum
associated with a charge-monopole pair.

The monopole analogues of the ordinary spherical harmonics were first
constructed by Tamm\Ref\tamm{I.~Tamm, Z.~Phys. {\bf 71}, 141 (1931).}
in the context of determining the wave function of an electron in the
field of a magnetic monopole.  The subject was revisited by Wu and
Yang\Ref\wuyang{T.T.~Wu and C.N.~Yang, Nucl. Phys. {\bf B107}, 365
(1976).}, whose conventions and notation I follow.
Olsen, Oslund, and Wu\Ref\olsen{H.A.~Olsen, P.~Osland,
and T.T.~Wu, Phys. Rev.~D{\bf 42}, 665 (1990).} obtained monopole
vector harmonics from the scalar harmonics by utilizing
Clebsch-Gordan technology.  In this paper I also study monopole
vector harmonics, but from a somewhat different approach.  Rather
than using Clebsch-Gordan coefficients, I construct the vector
harmonics (apart from an exceptional case described below) by applying
vector differential operators to the scalar harmonics, in analogy with
the construction often used\Ref\jackson{See, e.g., P.M.~Morse and
H.Feshbach, {\it Methods of Mathematical Physics} (McGraw-Hill, New
York, 1953), p.~1898.} for the ordinary ($q=0$) vector
harmonics.  Not only are the resulting expressions simpler, but they
are also more convenient for use in further calculations.  In addition,
with this approach the vector harmonics can be chosen to be
eigenfunctions of the radial component of the spin, rather than of the
magnitude of the orbital angular momentum, as was done in Ref.~\olsen.
With this choice of basis, the expressions for the curls and
divergences of the vector harmonics are easily derived and
take particularly simple forms.  There is also a
natural separation between radial and transverse vectors, making this
choice particularly useful for studying fields in spherically
symmetric  but curved spacetimes, such as one encounters when studying
fields about magnetically charged black holes.

I consider fields with electric charge $e$ in the presence of a
monopole with magnetic charge $q/e$; the Dirac quantization condition
restricts $q$ to integer or half-integer
values\rlap.\footnote{1}{Throughout this paper it will be assumed that
$q\ge 0$; the extension of the analysis to negative values of $q$ is
straightforward.} The corresponding scalar monopole spherical harmonics
$Y_{qLM}(\theta,\phi)$ are eigenfunctions of ${\bf L}^2$ and $L_z$,
where $$ {\bf L} = {\bf r}\times ({\bf p} - e{\bf A}) -q\hr
    \eqn\ldef$$
with $\hr = {\bf r}/r$.  The first term on the right hand side is the
usual orbital angular momentum, while the second is the extra
charge-monopole angular momentum.  These are orthogonal, so
classically $|{\bf L}| \ge q$.  Correspondingly, although the quantum
numbers $L$ and $M$ have their usual meaning, the minimum value of $L$
is not zero, but $q$.

     The monopole vector spherical harmonics are eigenfunctions of
${\bf J}^2$ and $J_z$, where the total angular momentum ${\bf J}$ is
the sum of $\bf L$ and the spin angular momentum ${\bf S}$.  By the
usual rules of for adding angular momenta, one sees that the total
angular momentum quantum number $J$ has a minimum value of $q-1$,
except for the two cases $q=0$ and $q=1/2$ (where $J_{\rm min}=q$).
The vector harmonics with $J\ge q$ can be constructed by applying
vector operators to the scalar harmonics. The vector harmonics with the
minimum allowed angular momentum, $J=q-1$, cannot be constructed in
this manner (essentially, because there
are no scalar harmonics with $L<q$) and so must be treated specially.
However, it turns out that these latter span the space of vectors whose
covariant curl and covariant divergence both vanish.  An index theorem
shows that this space has dimension $2q-1$, just as one would expect
for a multiplet with angular momentum $q-1$.

     The remainder of this paper is organized as follows.  In Sec.~2
the scalar monopole harmonics are reviewed and some general properties
of the vector harmonics are derived.  The construction of the vector
harmonics with $J \ge q$ is described in Sec.~3,
where some properties of these harmonics are derived.  The exceptional
case $J=q-1$ is discussed in Sec.~4.   The relationship between these
harmonics and those of Ref.~\olsen\ is given in an Appendix.

\chapter{General Considerations}

     In the absence of spin, the angular momentum operator in the
presence of a magnetic monopole may be written as
$$  {\bf L} = -i{\bf r}\times {\bf D} - q {\bf \hr}
        \eqno\eq $$
where $ {\bf D} = {\bf \nabla} - ie {\bf A} $ is the gauge
covariant derivative.   One can readily verify that $\bf L$ satisfies the
usual angular momentum commutation relation $[L_i,L_j] = i
\epsilon_{ijk}L_k$.  It also obeys the useful identity
$$  {\bf \tD}^2 = -{1\over r^2} \left( {\bf L}^2 -q^2\right)
      \eqn\anglap $$
where
$$ {\bf \tD} \equiv {\bf D} - \hr \hr \cdot { \bf D}
    \eqno\eq $$
is the purely angular part of the covariant derivative.

     The monopole spherical harmonics $Y_{qLM}(\theta,\phi)$ obey
$$ \eqalign{ {\bf L}^2 Y_{qLM} &= L(L+1) Y_{qLM} \cr
       L_z Y_{qLM} &= M Y_{qLM}  }
     \eqn\scaleq $$
where $L$ and $M$ can take on the values
$$  \eqalign{  L &= q, \, q+1, \,\dots \cr
               M& = -L,\, -L+1,\, \dots ,L }
     \eqno\eq $$

     When Eq.~\scaleq\ is solved to give an explicit expression for these
harmonics, one finds that they possess singularities that coincide
with the Dirac string\Ref\dirac{P.A.M.~Dirac, Proc. Roy.
Soc. {\bf A133}, 60 (1931).} of the monopole.  However, as was pointed
out by Wu and Yang\refmark\wuyang, the harmonics are in fact
nonsingular, provided that they are viewed as sections rather than as
ordinary functions.  In this approach, one divides the space outside
of the monopole into two overlapping regions.  For each region one
makes a choice of the vector potential, and thus of the monopole
harmonics, which is nonsingular within that region.  In the overlap of
the two regions the two vector potentials, and the two sets of
monopole harmonics, are related by a nonsingular gauge transformation
characterized by $q$.   The explicit form of the harmonics depends
on the choice of gauge and of the two regions.  Expressions for a
particularly convenient choice are given in Ref.~\wuyang, although we
will not need these in this paper.

    The $Y_{qLM}$ are orthonormal, with
$$ \int d\Omega\, Y_{qLM}^* Y_{qL'M'} \equiv \int_0^\pi d\theta
    \int_0^{2\pi}d \phi \,  Y_{qLM}^* Y_{qL'M'} = \delta_{LL'}\,\delta_{MM'}
     \eqn\scalarnorm$$
Further, they form a complete set, in that any smooth section can be
expanded as a linear combination of monopole harmonics.

     The monopole vector spherical harmonics are
eigenfunctions of ${\bf J}^2$ and $J_z$.  The allowed values of the
total angular momentum quantum number  $J$ are
$q-1$, $q$, $\dots$, except in the two cases $q=0$ and $q=1/2$ where
$J=q-1$ is absent. In general, there is more
than one way to obtain a given value of $J$, and thus several
multiplets of harmonics with the same total angular momentum. If, as
was done in Ref.~\olsen, the harmonics are chosen to be  eigenfunctions
of ${\bf L}^2$, as well as of ${\bf J}^2$ and $J_z$, the multiplet
structure is\footnote{2}{Note that a $J=0$ mode occurs only for $q=0$
and  $q=1$.  Thus, for any other value of $q$ it is impossible to
construct a spherically symmetric configuration involving a charged
vector field. In the context of a spontaneously broken $SU(2)$ gauge
theory, this explains why nonsingular spherically symmetric monopole
configurations are possible only for unit magnetic charge\Ref\guth{A.H.
Guth and E.J. Weinberg, Phys. Rev.~D{\bf 14}, 1660, 1976.}.}
\item{} For $J=q-1\ge 0$: one multiplet, with $L=J+1$.
\item{} For $J=q>0 $: two multiplets, with $L=J+1$ and $L=J$.
\item{} For $J=q=0 $: one  multiplet, with $L=1$.
\item{} For $J>q$: three multiplets, with $L=J+1$, $L=J$, and $L=J-1$.

     An alternative approach, which I follow in this paper, is to
classify the multiplets by
the eigenvalue of $ \bf\hr\cdot S $, which will be denoted $\lambda$.
In general, $\lambda$ can take on the values 1, 0, and $-1$.  However,
it is further restricted by the requirement that
$$ {\bf \hr\cdot J} = {\bf \hr\cdot L} + {\bf \hr\cdot S}
     = -q + \lambda
   \eqno\eq $$
lie in the range $-J$ to $J$.
This gives
\item{} For $J=q-1\ge 0$: one multiplet, with $\lambda = 1$.
\item{} For $J=q>0$: two multiplets, with  $\lambda = 1$ and 0.
\item{} For $J=q=0 $: one multiplet, with $\lambda=0$.
\item{} For $J>q$: three multiplets, with $\lambda = 1$, 0, and $-1$.

     Thus, let us denote the vector harmonics by
$\cl_{qJM}$, with
$$\eqalign{ {\bf J}^2\, \cl_{qJM} &= J(J+1) \,\cl_{qJM} \cr
    J_z \,\cl_{qJM} &= M \,\cl_{qJM} \cr
    {\bf \hr \cdot S} \,\cl_{qJM} &= \lambda \,\cl_{qJM} }
     \eqno \eq $$
Because the spin matrices $(S^k)_{ij} = -i \epsilon_{ijk}$, the last
of these equations is equivalent to
$$     \hr \times \cl_{qJM} = -i \lambda \cl_{qJM}
       \eqn\rcross $$
{}From this we see that the $\lambda=0$ harmonics are purely radial vectors,
while those with $\lambda=\pm 1$ are transverse.
It also follows that
$$  (\lambda' - \lambda)\,\cls_{qJM} \cdot \clp_{qJ'M'} =
      \left(i \hr \times \cls_{qJM}\right)  \cdot \clp_{qJ'M'}
    +i\cls_{qJM}  \cdot\left( \hr \times \clp_{qJ'M'}\right)
     = 0
    \eqno\eq $$
so that any two vector harmonics with different values of $\lambda$ are
orthogonal as vectors at every point.

   Furthermore, the usual methods can be used to show that any two
harmonics with different values of $J$, $M$, or $\lambda$ are
orthogonal in the functional sense.  It will be convenient to
normalize them so that
$$ \int d\Omega\, \cls_{qJM} \cdot \clp_{qJ'M'}
      = {\delta_{JJ'}\,\delta_{MM'}\,\delta_{\lambda\lambda'} \over r^2}
           \eqn\vectornorm $$
With this normalization, the vector harmonics are homogeneous of
degree $-1$ in the Cartesian coordinates, so that
$$   ({\bf r}\cdot {\bf D})\, \cl_{qJM} = - \cl_{qJM}
    \eqn\homog $$
In addition, Eqs.~\rcross\ and \vectornorm, together with the
observation that  ${\bf r}\cdot \cl_{qJM}$ is a scalar, imply that
$$ {\bf r}\cdot \cl_{qJM}  = \delta_{\lambda 0} Y_{qJM}
     \eqn\rdotC $$

\chapter{Harmonics for $J\ge q$}

      The ordinary vector harmonics can be constructed by
applying vector operators to the $Y_{LM}$.  In this section I
generalize this construction to obtain the
monopole vector harmonics for
$J\ge q$. To begin, let $\bf v$ be any vector operator
constructed from $\bf r$ and $\bf D$.  The commutation relation
$ [L_i, v_j] = i \, \epsilon_{ijk}\, v_k $  implies that
$$ \eqalign { [{\bf L}^2, v_k] &= -2i \epsilon_{ijk} L_i v_j - 2v_k \cr
          & = -2 ({\bf L\cdot S v})_k - 2 v_k  }
     \eqno \eq $$
Hence,
$$  \left( {\bf L} + {\bf S} \right)^2 {\bf v}Y_{qKM}
    = {\bf v}{\bf L}^2 Y_{qKM} = K(K+1) Y_{qKM}
    \eqno \eq $$
Thus if ${\bf v}_\lambda$ is a vector satisfying
$ \hr \times {\bf v}_\lambda = -i \lambda {\bf v}_\lambda $, then
the desired $ \cl_{qJM}$ will be given, up
to a (possibly $r$-dependent) normalization factor, by
${\bf v}_\lambda Y_{qJM}$.  A set of such vectors is
$$ \eqalign { {\bf v}_{\pm 1}  &= r {\bf D} \pm i{\bf r \times D} \cr
         {\bf v}_0 &= \hr}
    \eqno\eq $$

     The normalization factor for the harmonics with $\lambda=\pm 1$
can be  determined by using Eq.~\anglap, together with the fact that
$\bf D$ and $\bf \tD$ are equivalent when acting on the $Y_{qJM}$, to
obtain
$$  \eqalign {\int d\Omega \left|{\bf v}_{\pm 1} Y_{qJM}\right|^2
   &= \int d\Omega  \left|\left( r {\bf \tD} \pm i{\bf r}
     \times {\bf \tD}  \right)     Y_{qJM}\right|^2 \cr
   &= \int d\Omega Y_{qJM}^*  \left[ -2r^2 {\bf \tD}^2 \pm 2ir {\bf
      r\cdot \tD \times \tD}\right] Y_{qJM} \cr
   &= \int d\Omega   Y_{qJM}^*  \left[ 2({\bf L}^2-q^2) \pm 2er \,
      \epsilon_{ijk}\, r_i\, F_{jk} \right] Y_{qJM} \cr
   &= 2r^2 [{\cal J}^2  \pm q] }
   \eqn\normcalc$$
where
$${\cal J}^2 \equiv J(J+1)-q^2
      \eqno\eq$$
(Note that the integral in Eq.~\normcalc\ vanishes for $J=q>0$ and
$\lambda=-1$ and
for $J=q=0$ and $\lambda=\pm 1$, in accord with the earlier statement that
the  corresponding harmonics should be absent.)  For $\lambda=0$,
the normalization integral simply reduces to Eq.~\scalarnorm.
Thus, the properly normalized vector harmonics are
$$ \eqalign{ \c_{qJM}^{(1)} &=  \left[2 ({\cal J}^2 + q)\right]^{-1/2}
   \left[  {\bf D} + i{\hr \times \bf D}\right] Y_{qJM}
    ,\qquad \qquad  J\ge q >0\cr
      &\phantom{1} \cr
     \c_{qJM}^{(0)} &= {1\over r} \hr Y_{qJM}
 ,\qquad \qquad\qquad\qquad \qquad\qquad \qquad\quad\,\,\,\,J\ge q
\ge0\cr
      &\phantom{1} \cr
  \c_{qJM}^{(-1)} &= \left[2 ({\cal J}^2 - q)\right]^{-1/2}
   \left[  {\bf D} - i{\hr \times \bf  D}\right] Y_{qJM}
    ,\qquad \qquad  J> q \ge 0}
         \eqno\eq $$

    (One might think that it would have been simpler to choose two of the
vector harmonics to be proportional to ${\bf D}Y_{qJM}$ and
${\hr \times \bf  D}Y_{qJM}$, by analogy with the common practice in the
$q=0$ case\refmark\jackson. The problem is that these are not
orthogonal if $q\ne 0$; their orthogonality for $q=0$ follows from
parity arguments, but parity is not a good quantum number in the
presence of the monopole.)

  It is useful to have formulas for the covariant curls and divergences
of these vectors.  For $\lambda =0$,
$$ \eqalign {{\bf D} \times \c_{qJM}^{(0)}
     &= - {1\over r} \hr \times {\bf D} \,Y_{qJM}  \cr
     &= {i\over r }   \left[\sqrt{{\cal J}^2 +q\over 2}\,
              \c_{qJM}^{(1)}
      - \sqrt{{\cal J}^2 -q\over 2}\, \c_{qJM}^{(-1)} \right]  }
    \eqn\curlzero $$
and
$$ {\bf D}\cdot  \c_{qJM}^{(0)}
   = {\bf D} \cdot \left({\hr \over r}\right) Y_{qJM}
    = {1\over r^2} Y_{qJM}
 \eqn\divzero$$

    For $\lambda = \pm 1$, we first note that
$$   {\bf r} \times \left({\bf D} \times \c_{qJM}^{(\pm 1)}\right) =
    {\bf D} \left({\bf r}\cdot \c_{qJM}^{(\pm 1)}\right) -
               \c_{qJM}^{(\pm 1)}
    - ({\bf r}\cdot {\bf D}) \c_{qJM}^{(\pm 1)} = 0
   \eqn\curlcurl $$
where the vanishing of the first term on the right hand side
follows from Eq.~\rdotC, while
the cancellation of the last two terms is a consequence of
Eq.~\homog.  (This was the motivation for choosing the normalization condition
\vectornorm.)  Thus, the covariant curl of $\c_{qJM}^{(\pm 1)}$ is a
vector in the radial direction with magnitude $\hr \cdot {\bf D} \times
\c_{qJM}^{(\pm 1)}$.  The latter quantity is a scalar and can therefore
be expanded in scalar harmonics, with the coefficient functions
determined by the integrals
$$ \eqalign {\int d\Omega \, Y_{qJ'M'}^*\,\hr
    \cdot {\bf D} \times \c_{qJM}^{(\pm 1)}
   &= \int d\Omega \,Y_{qJ'M'}^*\hr \cdot {\bf \tD} \times \c_{qJM}^{(\pm 1)}
\cr
   &= r\int d\Omega \, \c_{qJ'M'}^{(0)*}
           \cdot {\bf \tD} \times \c_{qJM}^{(\pm 1)} \cr
   &= -r\int d\Omega \, {\bf \tD} \times\c_{qJ'M'}^{(0)*}
           \cdot\c_{qJM}^{(\pm 1)} \cr
   &= -r\int d\Omega \, {\bf D} \times\c_{qJ'M'}^{(0)*}
          \cdot\c_{qJM}^{(\pm 1)} \cr
   &= \pm {i\over r^2} \sqrt{ {\cal J}^2 \pm q\over 2} \,
       \delta _{JJ'}\, \delta_{MM'} }
    \eqno\eq $$
Hence
$$  {\bf D} \times \c_{qJM}^{(\pm 1)} =
    \pm {i\over r} \sqrt{ {\cal J}^2 \pm q\over 2} \, \c_{qJM}^{(0)}
    \eqn\onecurl $$

   The formula for the divergence is obtained by observing that
$$ \eqalign{
{\bf D}\cdot \c_{qJM}^{(\pm 1)} &= \pm i {\bf D}
                  \cdot \hr\times \c_{qJM}^{(\pm 1)}
    = \mp i \hr \cdot {\bf D} \times \c_{qJM}^{(\pm 1)} \cr
    &=  {1\over r^2} \sqrt{ {\cal J}^2 \pm q\over 2} \, Y_{qJM} }
     \eqn\onediv$$

    These results for the covariant divergences and curls can be
compactly summarized by
$$ {\bf D}\cdot \cl_{qJM} = {1\over r^2} a_\lambda Y_{qJM}
   \eqno\eq$$
with
$$ a_0 = 1, \qquad \qquad a_{\pm1} = \sqrt{{\cal J}^2 \pm q \over 2}
   \eqno\eq$$
and
$$ {\bf D}\times \cl_{qJM} = {i\over r} \sum_{\lambda'}
          b_{\lambda\lambda'}\clp_{qJM}
    \eqno\eq $$
where the only nonvanishing $b_{\lambda\lambda'}$ are  $b_{\pm 1,0} =
b_{0,\pm 1} = \pm a_{\pm1}$.  Furthermore, although they have only
been derived thus far for $J \ge q$, we will see in the next section that the
covariant curls and divergences of the $\c_{q(q-1)M}^{(1)}$  vanish,
in agreement with the above equations,
so that these expressions are in fact valid for all allowed values of
$J$, $M$, and $\lambda$.

\chapter{Vector harmonics for $J=q-1$}

\section{Curls and Divergences}

     For $J=q-1$ there is a single multiplet of vector harmonics, with
$L=q$ and $\lambda=1$.  These cannot be constructed by the methods of
the previous section, since there are no scalar harmonics for $J<q$.
In this section I first show that the covariant curls and divergences
of these harmonics vanish.  I then prove an index theorem showing that
the space of such curl-free and divergenceless vectors on the unit
two-sphere has dimension $2q-1$, and is thus spanned by the $J=q-1$
multiplet.  Finally, the harmonics are displayed explicitly for a
particular gauge choice of the vector potential. To simplify notation,
let $\c_{q(q-1)M}^{(1)} \equiv \u_M$.

     Consider first the divergence of $\u_M$.  Since this is a
scalar, it will vanish if
$$ I_{J'M'} \equiv \int d\Omega \, Y_{J'M'}^* {\bf D} \cdot \u_M
    \eqno\eq $$
vanishes for all possible values of $J'$ and $M'$.  To this end,
note that the fact that $\lambda=1$ implies that $\hr\cdot U_M=0$, from
which it follows that  ${\bf D}
\cdot \u_M = {\bf \tD} \cdot \u_M$.   Hence,
$$ \eqalign {I_{J'M'}
    &= - \int d\Omega \,(\tD Y_{J'M'})^*  \cdot \u_M \cr
    &=  - \int d\Omega
  \left[ \sqrt{{\cal J'}^2 +q\over 2} \, \c_{qJ'M'}^{(1)*} +
    \sqrt{{\cal J'}^2 -q\over 2} \, \c_{qJ'M'}^{(-1)*}  \right] \cdot
\u_M }
    \eqno\eq $$
But the last integral must vanish, since $J' \ge q$ while $\u_M$ is
a vector harmonic with angular momentum $J=q-1$.  Hence,
$$  {\bf D} \cdot \u_M = 0
      \eqn\divu $$
Proceeding as in Eq.~\onediv, we see that this imples this the vanishing
of $\hr \cdot {\bf D} \times \u_M $, while, from Eq.~\curlcurl,
$\hr \times ({\bf D} \times \u_M ) =0$.  Therefore
$${\bf D} \times \u_M =0
      \eqno\eq $$

     The $\u_M$ are thus a set of $2q-1$ linearly independent curl-free
and divergenceless vector fields.  Conversely, any vector field whose
covariant curl and divergence both vanish is a linear combination of
the $\u_M$.  To see this, expand the field in vector harmonics and
then use Eqs.~\curlzero, \divzero, \onecurl, and \onediv\ to show that
the coefficients of the $\cl_{qJM}$ vanish for $J\ge q$.

\section{An Index Theorem}

   Any vector field with vanishing curl and divergence is fixed
uniquely by its values on the unit two-sphere.  Since, in
addition, the $\u_M$ are orthogonal to $\hr$, they are equivalent to a
set of curl-free and divergenceless vector fields on this two-dimensional
manifold.   Let the metric on the two-sphere be denoted
$g_{ab}$, and the coordinate plus gauge covariant derivative be
$\calD_a$.  One can define a duality transformation, with the dual of
vector $V^a$ being
$$   \tilde V^a = -{i\over \sqrt{g}} \epsilon^{ab} V_b
       \eqno \eq $$
where $g\equiv \det\,g_{ab}$ and $\epsilon^{ab}$ is the antisymmetric
symbol with
$\epsilon^{12} =\epsilon^{\theta\phi} = 1$.
Three-dimensional vectors with $\lambda = 1$ ($\lambda = -1$)
correspond to self-dual (anti-self-dual) vectors on the two-sphere.
The operators $P_+$ ($P_-$) projecting onto the space of self-dual
(anti-self-dual) vectors are
$$  P^{ab}_\pm  = {1\over 2} \left[ g^{ab}
       \mp {i\over \sqrt{g}}\, \epsilon^{ab} \right]
     \eqno\eq $$
With the aid of the identity $\epsilon ^{ab} \epsilon_{bc} = - g^a_c\,
g$, one can verify that $P_\pm^a{}_b  P_\pm^b{}_c = P_\pm^a{}_c$.
Furthermore, $\calD_a P_\pm^{bc} =0$.

   The space of curl-free self-dual (anti-self-dual) vectors may be
identified as the  kernal of the
operator $\calO_+$  ($\calO_-$) mapping such vector fields onto the
space of scalar fields according to
$$  \eqalign {  \calO_\pm V
   &= {1\over \sqrt{g}} \,\epsilon^{ab} \calD_a  P_{\pm bc}V^c \cr
    &= \mp i \calD_a P_\pm^a{}_b V^b }
      \eqno\eq$$
The second equality shows that any curl-free self-dual or
anti-self-dual vector must also be divergenceless.   Conversely, it is easy
to see that any
vector with vanishing curl and divergence must be either self-dual or
anti-self-dual.

    Angular momentum  considerations suggest
that the dimension of the kernal of $\calO_+$
should be $2q-1$ for $q>0$, and that the kernal should vanish for
$q=0$.  Furthermore, since an anti-self-dual curl-free vector field
would correspond to a field with the forbidden values $\lambda=-1$ and
$J=q-1$, the kernal of $\calO_-$ should vanish.
These results correspond to index theorems relating the magnetic charge to
the index
$$ {\cal I}(\calO_\pm) = {\rm dim(kernal}~\calO_\pm)
      -{\rm dim(kernal}~ \calO_\pm^\dagger)
    \eqno\eq $$
Here the adjoint operators, $\calO_\pm^\dagger$,
mapping scalar fields onto self-dual or anti-self-dual vector fields, are
$$   (\calO_\pm^\dagger S)^a = -i (P_\pm^{ba})^* \calD_b  S
          = -i P_\pm^{ab} \calD_b  S
      \eqno\eq$$

     The first step in deriving these theorems is to note that
the kernals of $\calO_\pm$ and $\calO_\pm^\dagger$ are the same as
those of $\calO_\pm^\dagger\calO_\pm$ and
$\calO_\pm\calO_\pm^\dagger$, respectively.  Furthermore,
if $\psi$ is an eigenfunction of $\calO_\pm^\dagger\calO_\pm$ with
nonzero eigenvalue, then $\calO_\pm \psi$ is an eigenfunction of
$\calO_\pm\calO_\pm^\dagger$ with the same eigenvalue.  Assuming that
these eigenfunctions form a complete basis, it follows that
$$ {\cal I}(\calO_\pm) =
  {\rm Tr}  \left({M^2 \over 2\calO_\pm^\dagger\calO_\pm + M^2} \right)
 -   {\rm Tr}  \left({ M^2 \over 2\calO_\pm\calO_\pm^\dagger + M^2} \right)
    \eqn\indexexp $$
where $M^2$ is an arbitrary parameter.  (The somewhat unconventional
factors of 2 are for later convenience.)
 We will find it most convenient to
evaluate this expression in the limit $M^2 \rightarrow \infty$.

     With the aid of the identities $P_\pm^2=P_\pm$ and
$$ P_\pm^{ab}P_\pm^{cd} = P_\pm^{ad} P_\pm^{cb}
     \eqno\eq $$
one finds that
$$  \eqalign{  (\calO_\pm^\dagger\calO_\pm)^a{}_b &=
    - P_\pm^a{}_b  P_\pm^{cd}\calD_d \calD_c \cr
   &= -{1\over 2}P_\pm^a{}_b  \left( \calD^c \calD_c
   \pm{i \over 2\sqrt{g}}\, \epsilon^{cd}\, [\calD_c, \calD_d] \right) }
     \eqno\eq $$
and
$$  \eqalign{ \calO_\pm\calO_\pm^\dagger &= - P_\pm^{cd}\calD_c \calD_d  \cr
     &= -{1\over 2}  \left( \calD^c \calD_c
   \mp {i \over 2\sqrt{g}}\, \epsilon^{cd}\, [\calD_c, \calD_d] \right) }
     \eqno\eq $$
The commutator of two coordinate and gauge covariant derivatives
$\calD_a$ is the sum of the commutator of the corresponding gauge
covariant derivatives $D_a$ and the commutator of the corresponding
coordinate covariant derivatives $\nabla_a$.  Acting on charged
fields, the former is
$$  [D_a, D_b] = -ie F_{ab} = -{iq\over\sqrt{g}}\, \epsilon_{ab}
    \eqno\eq $$
where the second equality follows from the expression for the
magnetic field at unit radius from the monopole. The latter
commutator vanishes when acting on scalar fields, while on
self-dual or anti-self-dual vector fields
$$  \eqalign{ \epsilon^{cd} \,[\nabla_c, \nabla_d] V^a
   &= \epsilon^{cd}\, [\nabla_c, \nabla_d] P_\pm^a{}_bV^b  \cr
   &= - \epsilon^{cd} P_\pm^a{}_b R^b{}_{ecd} V^e \cr
   &=  \epsilon^{cd} P_\pm^a{}_b \left( g^b{}_c\, g_{ed}- g^b{}_d\, g_{ec}
         \right) V^e \cr
   &= \pm 2i \sqrt{g} \,P_\pm^a{}_e V^e \cr
   &= \pm 2i \sqrt{g} \, V^a  }
     \eqno\eq $$
where the explicit form of the
curvature tensor on the two-sphere has been used on the third line.
We thus have
$$  (\calO_\pm^\dagger\calO_\pm)^a{}_b = {1\over 2}P_\pm^a{}_b
       \left( - \calD^c \calD_c \mp q +1 \right)
      \eqn\ohadjoh $$
and
$$  \calO_\pm\calO_\pm^\dagger = {1\over 2}
       \left( - \calD^c \calD_c \pm q  \right)
      \eqn\ohohadj $$
The factors of $q$ enter with the opposite sign for $\calO_+$ and
$\calO_-$ because the magnetic field is parity-violating and so
distinguishes between self-dual and anti-self-dual fields.
There is no such asymmetry as far as the geometry of the sphere is
concerned, and so the factor of 1 arising from the
curvature displays no sign change.

    Eq.~\ohadjoh\ shows that $\calO_-^\dagger\calO_-$ is positive definite
for $q\ge 0$.  Hence, the kernal of $\calO_-^\dagger$ vanishes, and
there are no curl-free anti-self-dual vector
fields, as expected.  Furthermore, since $ \calD^c \calD_c$ is
equivalent to  $-({\bf L}^2 - q^2)$ when acting on scalar fields,
Eq.~\ohohadj\ may be rewritten as
$$  \calO_\pm\calO_\pm^\dagger = {1\over 2}
       \left[ ({\bf L}^2 -q^2) \pm q  \right]
      \eqno\eq $$
The eigenfunctions of $\calO_\pm\calO_\pm^\dagger$ are thus the scalar
monopole harmonics $Y_{qLM}$, with eigenvalues $L(L+1)-q(q\mp 1)$.
For the lower signs, a zero eigenvalue is
obtained only for $L=q$.  Since there are $2L+1=2q+1$ possible values of
$M$, the kernal of $\calO_-^\dagger$ has dimension $2q+1$, and
$$ {\cal I}(\calO_-) = -2q-1
    \eqn\minusindex$$
For the upper signs, a zero eigenvalue is possible only
for $q=0$, $L=0$.  Thus, for $q>0$, the kernal of $\calO_+^\dagger$
vanishes, and the dimension of the
kernal of $\calO_+$ is equal to ${\cal I}(\calO_+)$.

    To calculate this last quantity, substitute Eqs.~\ohadjoh\ and
\ohohadj\  into
Eq.~\indexexp, and expand the denominators about $-\calD^c \calD_c +M^2$.
Thus,
$$\eqalign { {\cal I}(\calO_\pm) &=
  M^2 {\rm Tr}  \left[{P_\pm \over -\calD^c \calD_c + M^2}
     \pm{( q \mp1) P_\pm  \over \left(-\calD^c \calD_c + M^2\right)^2 }
     + \cdots \right] \cr
&\qquad  -  M^2 {\rm Tr}  \left[{ 1 \over -\calD^c \calD_c + M^2}
     \mp{ q  \over \left(-\calD^c \calD_c + M^2\right)^2 }
     + \cdots \right] }
    \eqno\eq $$
where the dots represent terms which vanish in the limit $M^2
\rightarrow \infty$.  The contributions from the first terms in the two
expansions cancel\rlap.\footnote{3}{This is less obvious than it
might seem, since
the operator $ -\calD^c \calD_c$ is understood to be acting on vectors
fields in one case and scalar fields in the other, and so in curved
space will take two different forms.  However, one can
verify the cancellation by comparing the result for ${\cal I}(\calO_-)$
obtained below with that given in Eq.~\minusindex.}   To deal with the
second terms in the expansions, note that replacing $\calD^c \calD_c$
by the flat-space two-dimensional Laplacian $\Delta$ gives an error of
order $M^{-2}$.  Therefore
$$\eqalign{ {\cal I}(\calO_\pm)
     &= (\pm 2q - 1) \lim_{M^2 \rightarrow \infty} \int  d^2x
    M^2 \langle x| (- \Delta + M^2)^{-2} |x\rangle \cr
    &= (\pm 2q - 1) \lim_{M^2 \rightarrow \infty} \int  d^2x
    \int {d^2k \over (2\pi)^2} {M^2 \over (k^2 +M^2)^2 } \cr
    &= \pm 2q -1 }
    \eqno\eq $$

\section{Explicit Expressions}

    To obtain  explicit expressions for the ${\bf U}_M$, we must
first fix the vector potential.  Following Wu and Yang\refmark\wuyang,
let $R_a$ be the region $0 \le \theta <  (\pi/2) + \delta$ and $R_b$ be
the region $(\pi/2)- \delta < \theta \le \pi$.  A nonsingular choice
for the vector potential which maintains the manifest rotational
symmetry about the $z$-axis is
$$ \eqalign {{\bf A}_\phi &= {q\over e} ( 1 - \cos\theta), \qquad
              {\rm in}~R_a \cr
      {\bf A}_\phi &= -{q\over e} (1 +  \cos\theta), \qquad
               {\rm in}~R_b }
    \eqno\eq $$
The action of $J_z$ on a scalar function is then
$$ J_z \psi = L_z \psi= \left( -i \partial_\phi \mp q \right ) \psi
     \eqno\eq $$
where the upper (lower) sign refers to region $R_a$ ($R_b$).
Applying this to the scalar ${\bf \hat z} \cdot {\bf U}_M$, and using
the fact that $\bf \hat z$ is invariant under rotations about the
$z$-axis, we find that the eigenvalue equation $J_z{\bf U}_M = M{\bf
U}_M$ implies
$$  \left( -i \partial_\phi \mp q \right ) ({\bf U}_M)_\theta
     = M ({\bf U}_M)_\theta
     \eqn\eigen$$
Hence  $({\bf U}_M)_\theta$ must be of the form
$$  ({\bf U}_M)_\theta = e^{i(M \pm q)\phi} f_{qM}(\theta)
     \eqn\thetacomp$$
The self-duality condition then gives
$$ ({\bf U}_M)_\phi
    = i\,e^{i(M \pm q)\phi} \sin\theta \, f_{qM}(\theta)
     \eqno\eq $$
Substituting Eq.~\thetacomp\ into Eq.~\divu, we obtain
$$ \partial_\theta (\sin\theta \,f_{qM}) -(M + q\cos \theta) f_{qM} =0
     \eqno\eq $$
whose solution is
$$ \eqalign {f_{qM}(\theta) &=  a_{qM}
     \left[{1-\cos\theta \over 1+ \cos\theta}\right]^{ M/2}
        \sin^{q-1}\theta \cr
    &=  a_{qM}
     (1-\cos\theta)^M
        \sin^{q-M-1}\theta \cr
    &=  a_{qM}
     ( 1+ \cos\theta)^{-M}
        \sin^{q+M-1}\theta }
   \eqno\eq $$
The normalization constant $a_{qM}$ is determined (up to an arbitrary
phase) by Eq.~\vectornorm\ to be
$$  a_{qM} = { 1 \over 2^q \sqrt{2\pi} }
    \left[{ (2q-1)! \over (q+M-1)! (q-M-1)! }\right] ^{1/2}
      \eqno\eq $$

\def\parc{\left(\c_{q(q-1)M}^{(1)}\right)}

    Thus, the ${\bf U}_M$ may be written as
$$ \eqalign{ &\eqalign{ ({\bf U}_M)_\theta = \parc_\theta
   &= {a_{qM}\over r} e^{i(M + q)\phi}\sin^{q+M-1}\theta \,
     ( 1+ \cos\theta)^{-M}  \cr
    ({\bf U}_M)_\phi = \parc_\phi
   &={i a_{qM}\over r}  e^{i(M + q)\phi}\sin^{q+M}\theta \,
     ( 1+ \cos\theta)^{-M} } \qquad\,\,   {\rm in}~R_a \cr
    &\phantom{1} \cr
   &\eqalign{ ({\bf U}_M)_\theta =\parc_\theta
   &= {a_{qM}\over r} e^{i(M - q)\phi}\sin^{q-M-1}\theta \,
     ( 1- \cos\theta)^{M}  \cr
    ({\bf U}_M)_\phi = \parc_\phi
   &={i a_{qM}\over r}  e^{i(M -q)\phi}\sin^{q-M}\theta \,
     ( 1- \cos\theta)^{M} } \qquad\quad   {\rm in}~R_b }
      \eqn\umresult  $$
For these expressions to be single-valued, $q-M$ and $q+M$ must be
integers.  To avoid a singularity at $\theta =0$ (in region~$R_a$), we
must require $q+M-1 \ge 0$, while at $\theta=\pi$ (in region~$R_b$) we
have the condition $q - M-1 \ge 0$.  This leaves $2q-1$ allowed values
of $M$, thus confirming the index calculation and giving the full
$J=q-1$ angular momentum multiplet.

\ack I would like to thank Kimyeong Lee, Parameswaran Nair, and Alexander
Ridgway for helpful comments.  I am also grateful to the Aspen Center
for Physics, where part of this work was done.

\appendix

Olsen et al\refmark\olsen\ define monopole vector harmonics $\y$ which
are eigenfunctions of ${\bf J}^2$, ${\bf L}^2$, and $J_z$.  In this
appendix I obtain the relationship between these harmonics and the
$\cl_{qJM}$ defined in this paper.
Because
the $\y$ are orthonormal, while the $\cl_{qJM}$ are normalized according to
Eq.~\vectornorm, the two sets of harmonics are related by
$$  \cl_{qJM} = {1\over r} \sum_L M_{\lambda L}(q,J, M) \y
     \eqn\translation $$
where the matrices $M_{\lambda L}(q,J, M)$ are unitary.

   For $J=q-1$, there is only one allowed value each for $\lambda$ and
for $L$, and so $M(q, q-1,M)$ is simply a complex number of unit
magnitude.  Carrying out explicitly the construction of Ref.~\olsen\
and comparing with Eq.~\umresult, one finds that
$$  M(q, q-1,M) = (-1)^{q+M}
    \eqno\eq$$

   For larger values of $J$ the $M_{\lambda L}(q,J, M)$ are either
$2\times 2$ (if $J=q$) or $3\times 3$ (if $J>q$).  The $0L$ elements
of $M_{\lambda L}(q,J, M)$ can be obtained directly from Eq.~(4.13) of
Ref.~\olsen, which expresses $\hr Y_{qLM}$ in terms of the $\y$.
Explicitly,
$$  \eqalign{  M_{0(J-1)}(qJM) &= \sqrt{J^2-q^2 \over (2J+1)J} \cr
        M_{0J}(qJM)&= - {q\over \sqrt{J(J+1)}} \cr
      M_{0(J+1)}(qJM) &= -\sqrt{(J+1)^2-q^2 \over (2J+1)(J+1)}  }
      \eqno\eq $$

    The first step in obtaining the remaining elements of $M_{\lambda
L}(q,J, M)$ is to use Eq.~(4.5) of Ref.~\olsen,
$$ {\bf r} \times {\bf Y}^{(q)}_{JlM} = ir \sum_L A^{(q)}_{JlL}\y
     \eqn\rcrossolsen $$
where the matrices $A^{(q)}_{JlL}$ are given explicitly in
Ref.~\olsen.  Together with Eqs.~\rcross and \translation, this leads to
$$ \sum_l M_{\lambda l} A_{lL} = - \lambda M_{\lambda L}
    \eqno\eq $$
This, together with the unitarity of $M_{\lambda L}$, determines the
$M_{\lambda (J\pm1)}$ in terms of the $M_{\lambda J}$, and fixes the
latter up to a $\lambda$-dependent phase.  Specifically,
$$ \eqalign{M_{1J}(qJM)
    &= e^{i\alpha} \sqrt{{\cal J}^2 +q \over 2J(J+1)}  \cr
      M_{-1J}(qJM)
    &= e^{i\beta} \sqrt{{\cal J}^2 -q \over 2J(J+1)}  }
     \eqn\phaseeq$$
The next step is to use Eq.~(4.14) of Ref.~\olsen,
$$ {\bf L} Y_{qJM} = \sqrt{J(J+1)} \y
    \eqno\eq $$
Substituting Eqs.~\translation, \rcrossolsen, and \phaseeq\ into this and
equating coefficients of ${\bf Y}^{(q)}_{JJM}$ yields
$$ e^{i\alpha} ({\cal J}^2 +q) - e^{i\beta} ({\cal J}^2 -q )
    + 2{\cal J}^2 =0
     \eqno\eq $$
whose only solution is $e^{i\alpha} = -e^{i\beta} = -1$.  Hence,
$$ M_{\pm1J}(qJM) = \pm \sqrt{{\cal J}^2 \pm q \over 2J(J+1)}
    \eqno \eq $$
from which one obtains
$$ \eqalign{ M_{\pm1(J+1)}(qJM) &= \sqrt{{\cal J}^2 \pm q \over 4J(J+1)}
    \sqrt{J+1\pm q \over (J+1)(J+1 \mp q)}  \cr
    M_{\pm1(J-1)}(qJM) &= \sqrt{{\cal J}^2 \pm q \over 4J(J+1)}
    \sqrt{J\mp q \over J(J \pm q)}  }
   \eqno\eq $$
Note that if $J=q$, both $M_{-1L}$ and $M_{\lambda(J-1)}$ vanish, so
that $M_{\lambda L}$ is actually a $2\times 2$ matrix, as required.

\refout

\end